\documentclass[12pt]{article}
\usepackage{mathtools,amssymb,amsfonts,graphicx}

\usepackage{hyperref}
\usepackage[nosort]{cite}
\usepackage{verbatim}
\usepackage{multicol,color,xcolor,longtable}

\definecolor{darkred}{rgb}{0.65,0.15,0}
\definecolor{darkgreen}{rgb}{.05,.5,.05}
\hypersetup{pdfborder={0 0 0},colorlinks=true,urlcolor=darkred,citecolor=blue,linkcolor=darkred,linktocpage=true}

\newcommand{\nn}{\nonumber}

\newcommand{\mf}[1]{{\mathfrak{#1}}}
\newcommand{\Tr}{\text{Tr}}
\newcommand{\lb}{\left\{}
\newcommand{\rb}{\right\}}
\newcommand{\PB}{{\text{P.B.}}}
\newcommand{\DB}{{\text{D.B.}}}

\makeatletter

\@addtoreset{equation}{section}
\makeatother

\newcommand{\be}{\begin{equation}}
\newcommand{\ee}{\end{equation}}
\newcommand{\uo}{\underline{\mathbf{1}}}
\newcommand{\ut}{\underline{\mathbf{2}}}
\newcommand{\CCo}{\overset{\scriptscriptstyle{(1)}}C_{\underline{\mathbf{1}}\underline{\mathbf{2}}}}
\newcommand{\CCt}{\overset{\scriptscriptstyle{(0)}}C_{\underline{\mathbf{1}}\underline{\mathbf{2}}}}

\begin{document}

\mbox{}

\vspace{20mm}

\begin{center}
  {\LARGE \sc  Integrable auxiliary field\\[4mm] deformations of coset models}
    \\[13mm]

{\large
Mattia Ces\`aro${}^{1}$, Axel Kleinschmidt${}^{1,2}$ and  David Osten${}^3$}

\vspace{8mm}
${}^1${\it Max-Planck-Institut f\"{u}r Gravitationsphysik (Albert-Einstein-Institut)\\
Am M\"{u}hlenberg 1, DE-14476 Potsdam, Germany}
\vskip 1.2 ex
${}^2${\it International Solvay Institutes\\
ULB-Campus Plaine CP231, BE-1050 Brussels, Belgium}
\vskip 1.2 ex
${}^3${\it Institute for Theoretical Physics (IFT), University of Wroc\l aw,\\
pl. Maxa Borna 9, 50-204 Wroc\l aw, Poland}

\end{center}

\vspace{25mm}

\begin{center} 
\hrule

\vspace{6mm}

\begin{tabular}{p{14cm}}
{\small%
We prove the existence of a family of integrable deformations of $\mathbb{Z}_N$-coset models in two dimensions. Our approach uses and generalises the method of auxiliary fields that was recently introduced for the principal chiral model by Ferko and Smith. 
}
\end{tabular}
\vspace{5mm}
\hrule
\end{center}

\hypersetup{pageanchor=false}
\thispagestyle{empty}

\newpage
\hypersetup{pageanchor=true}
\setcounter{page}{1}


\tableofcontents

\section{Introduction and summary}
Non-linear $\sigma$-models are ubiquitous in various branches of physics. Their origin lies in particle physics as an effective theory of the strong interaction between pions \cite{GellMann:1960np}. However, their applicability turned out to be vastly broader with appearances in statistical and condensed matter physics, for example as models for random matrix ensembles \cite{Zirnbauer:1996zz} or disordered metals \cite{Efetov:1983xg}. They also find important applications in gravity \cite{Geroch:1970nt,Mazur:1983vi} and string theory. 

Remarkably, despite being non-linear, interacting quantum field theories, in two dimensions they can be classically integrable, with prime examples being the principal chiral model or the symmetric space $\sigma$-model. In the latter case, the theory is given by maps from the two-dimensional `world-sheet' to a target space $G/K$ where $G$ is a Lie group and $K$ a subgroup fixed by an involution of $G$, for example $K$ can be the maximal compact subgroup of $G$. While the integrability of symmetric space models has been known for some time \cite{Eichenherr:1979ci}, it is interesting to consider whether there are further, related integrable models. These are often referred to as integrable deformations. So far, the focus in the literature lay on deformations inside the class of $\sigma$-models, most prominently so-called Yang--Baxter \cite{Klimcik:2002zj,Klimcik:2008eq} and $\lambda$-deformations \cite{Sfetsos:2013wia} have been studied in the literature (see \cite{Hoare:2021dix} for a recent review). 

Recently, another type of deformations has been introduced for principal chiral models~\cite{Ferko:2024ali,Bielli:2024khq}. These deformations are obtained by introducing an auxiliary field and adding an arbitrary function of an invariant built from it to the Lagrangian. In comparison with the above mentioned examples, one can identify two main characteristics: First, these deformations leave the realm of $\sigma$-models since they introduce additional fields. Second,  they can still be considered to be quite mild deformations of $\sigma$-models, as they still they share some features. On the level of classical integrability, the latter can be understood by the fact that the original and deformed level both have the same twist function (see~\cite{Lacroix:2018njs} for a review on integrable models with twist function).

The form of the auxiliary field deformation~\cite{Ferko:2024ali} is inspired by reformulations of Max\-well's theory~\cite{Ivanov:2001ec,Ivanov:2002ab,Ferko:2023wyi} and the deformations were shown to be related to $T\bar{T}$- and root-$T\bar{T}$-\linebreak deformations~\cite{Zamolodchikov:2004ce,Cavaglia:2016oda,Babaei-Aghbolagh:2022uij,Ferko:2022cix,Borsato:2022tmu}. Similar considerations starting from the Yang--Baxter deformed principal chiral model~\cite{Bielli:2024fnp} and interpretation as deformations by higher-spin currents~\cite{Bielli:2024ach} have been provided as well. The auxiliary field deformation has also been embedded into the framework of 4d Chern--Simons-theory~\cite{Costello:2017dso} in~\cite{Fukushima:2024nxm}, where also the generalisation of the deformation to the WZW-model has been presented. 

In this article, we will extend this method from principal chiral models to coset $\sigma$-models, where we consider both the symmetric space cosets (with an underlying $\mathbb{Z}_2$-automorphism) but also of their $\mathbb{Z}_N$-generalisation, leading to $\mathbb{Z}_N$-coset spaces. We will obtain an infinite family of integrable models in this way. Integrable $\mathbb{Z}_N$-coset $\sigma$-models have been introduced in generality in \cite{Young:2005jv}. Their prime example has been introduced before and is that of the superstring $\sigma$-models on supergroup cosets associated to a $\mathbb{Z}_4$-grading, most famously in their Green--Schwarz formulation in flat space~\cite{Henneaux:1984mh} and AdS$_5 \times $S$^5$~\cite{Metsaev:1998it}, or in the pure spinor formalism~\cite{Berkovits:1999zq}. Interpretations of $\sigma$-models with $\mathbb{Z}_N$-coset target spaces with $N \neq 2,4$ have been given in~\cite{Osten:2021opf}. In any case, it will be interesting to understand the physical interpretation of the novel auxiliary field deformations found in this paper in full generality.

We also present a generalisation of the method of~\cite{Ferko:2024ali} by not starting from a Lagrangian deformation, but by considering a deformation of the Lax formulation of the dynamics which is less constraining and of which the Lagrangian deformation in~\cite{Ferko:2024ali}, and its straightforward generalisations for coset spaces considered here, is a special case.

The structure of this article is as follows. We first review symmetric spaces and their integrable structure in section~\ref{sec:rev} before introducing Lagrangian deformations and showing their integrability in section~\ref{sec:Z2def}. In section~\ref{sec:Zn}, we consider the case of $\mathbb{Z}_N$-gradings and integrable deformations that do not necessarily arise from Lagrangian deformations.

\medskip

\textbf{Note added.} On the same day as this work, the related paper \cite{Bielli:2024oif} also appeared on the arXiv. It contains partially overlapping but also complementary results to this work. The paper \cite{Bielli:2024oif} extends the higher-spin auxiliary field deformations of \cite{Bielli:2024ach} (more general than the spin-2 deformations considered in this work) to the case of $\sigma$-models for $\mathbb{Z}_2$- and $\mathbb{Z}_4$-cosets (less general than $\mathbb{Z}_N$ models with arbitrary $N$ considered in this work). Furthermore, the focus  of \cite{Bielli:2024ach} lies in the analysis of flat Lax representations (weak integrability) of higher-spin models, whereas we prove Hamiltonian (or strong) integrability for spin-2 deformations in the present work.

\section{Review of integrable symmetric space models}
\label{sec:rev}

We begin by recalling some features of integrable symmetric space models, including their Lax representation and Hamiltonian integrability. This material is discussed at various places in the literature and we rely mainly on~\cite{Lacroix:2018njs,Hoare:2021dix,Driezen:2021cpd} and follow~\cite{Matschull:1994vi} for the canonical formulation. 
In this and the next section we consider the symmetric space case, with a $\mathbb{Z}_2$-grading, and leave the more general case to section~\ref{sec:Zn}.

\subsection{Symmetric space sigma models}

For a symmetric space $G/K$ the global isometry $G$ has a Lie algebra $\mathfrak{g}$ that decomposes as
\begin{align}
\label{eq:kp}
\mathfrak{g}= \mathfrak{g}^{(0)} \oplus \mathfrak{g}^{(1)} = \mathfrak{k} \oplus  \mathfrak{p}\, .
\end{align}
This decomposition is a $\mathbb{Z}_2$-grading of $\mathfrak{g}$ and $\mathfrak{k}$ is the Lie algebra of a (connected) subgroup $K\subset G$, fixed by an involution. If $K$ is the maximal compact subgroup of $G$ then the coset $G/K$ is a Riemannian symmetric space, but the set-up is more general. We assume $G$ to be simple and write the invariant bilinear form on its Lie algebra by $\Tr$ and the Killing form is block-diagonal in this grading.

The coset $\sigma$-model depends on a field $\mathbb{R}^{1,1} \to G/K$, where we refer to $\mathbb{R}^{1,1}$ as (flat) space-time.\footnote{In string theory interpretations, this could be replaced a two-dimensional surfaces of signature $(-+)$, the world-sheet.}
We write the coset using a `vielbein' $\mathcal{V}(x)\in G$ 
that transforms as
\begin{align}
\label{eq:costrm}
\mathcal{V}(x) \to \mathcal{V}^\prime(x) =g^{-1}\mathcal{V}(x) k(x)\, ,
\end{align}
with global $g\in G$ and local (in space-time) $k(x) \in K$, corresponding to the freedom of choosing a coset representative of $G/K$ at each space-time point. Fixing a specific form of a representative for $G/K$  determines $k(x)$ as a compensating transformation that depends on $g$ and $\mathcal{V}(x)$. In the following we will mostly suppress the dependence on the space-time coordinate $x^\alpha$.

The  $\mathfrak{g}$-valued Maurer--Cartan current decomposes  according to~\eqref{eq:kp} as
\begin{align}
\label{eq:QP}
\mathcal{V}^{-1}\partial_\alpha  \mathcal{V}=P_\alpha +Q_\alpha\,, \quad P_\alpha\in\mathfrak{p}\, ,\; Q_\alpha\in\mathfrak{k} .
\end{align}
The two components transform covariantly and as a connection, respectively, under~\eqref{eq:costrm}:
\begin{align}
\label{eq:QPtrm}
P_\alpha \to  k^{-1}P_\alpha k\,, \quad
Q_\alpha  \to k^{-1}Q_\alpha k+ k^{-1}\partial_\alpha k\,.
\end{align}
From the definition~\eqref{eq:QP} one can deduce the Bianchi identities
\begin{align}
\label{eq:Bianchis}
2\partial_{[\alpha}Q_{\beta]}+[Q_\alpha,Q_\beta]=-[P_\alpha,P_\beta]\, , \hspace{15mm} 
D_{[\alpha}P_{\beta]}=0\, ,
\end{align}
where we have introduced the $K$-covariant derivative $D_\alpha = \partial_\alpha +Q_\alpha$.
Note that for a fixed choice of representative, the connection $Q_\alpha$ becomes composite. 

The flat space coset model can be defined by the Lagrangian
\begin{align}
\label{eq:undeformedLagrangianP}
\mathcal{L}=-\frac{1}{2}\eta^{\alpha\beta}\Tr (P_\alpha P_\beta)\, .
\end{align}
It is manifestly invariant under~\eqref{eq:costrm} in view of~\eqref{eq:QPtrm}. The equation of motion that follows from it by variation is 
\begin{align}
\label{eq:eomP}
D_\alpha P^\alpha = 0 \,,
\end{align}
where the index has been raised with the flat space-time metric $\eta^{\alpha\beta}$. 

\subsection{Lax representation of undeformed coset model}

It is well known that the equations of motion~\eqref{eq:eomP} can be obtained from a linear system. The compatibility of the linear system is equivalent to the dynamical equations and here can be taken as
\begin{align}
\label{eq:cosetLaxundeformed}
\mathfrak{L}_\alpha \equiv \hat{\mathcal{V}}^{-1}(z) \partial_\alpha \hat{\mathcal{V}}(z) =  Q_\alpha + \frac{1+z^2}{1-z^2} P_\alpha + \frac{2z}{1-z^2} \epsilon_{\alpha\beta} P^\beta\, ,
\end{align}
where $\epsilon_{\alpha\beta}$ is the Levi--Civita tensor that we take with $\epsilon^{01}=+1$. The new parameter $z\in \mathbb{C}$ is the spectral parameter and we have introduced a new object $\hat{\mathcal{V}}(z)$ that can be thought of as an element of the loop group of $G$ for suitable analytic conditions on the $z$-dependence. We will not use this aspect and simply view it as a generating series around $z=0$. 

The compatibility condition
\begin{align}
2\partial_{[\alpha}\mathfrak{L}_{\beta]}+[\mathfrak{L}_\alpha, \mathfrak{L}_\beta]=0
\end{align}
is equivalent to the Bianchi identities \eqref{eq:Bianchis}  as well as the equation of motion~\eqref{eq:eomP}. This is the statement that the coset model has a Lax representation.

\subsection{Hamiltonian integrability}
\label{sec:HamZ2}

Hamiltonian integrability can be shown by bringing  the system into Maillet form. This requires using splitting space and time derivatives on the two-dimensional space-time, which we denote by $x^\alpha=(t,x)$. 

The Maillet form~\cite{Maillet:1985ec,Maillet:1986} requires proving that the equal-time canonical bracket of the spatial part $\mathfrak{L}_x$ of the Lie-algebra valued Lax connection takes the special form 
\begin{align}
\label{eq:Maillet}
    \lb \mathfrak{L}_{x,\uo}(x,z), \mathfrak{L}_{x,\ut}(x',z')\rb 
    &= 
    \left[ r_{\uo\ut}(z,z^\prime),\mathfrak{L}_{x, \uo}(x,z)\right]
    -\left[r_{\ut\uo}(z^\prime, z)\mathfrak{L}_{x, \ut}(x',z^\prime)\right]
    \nn\\
&\quad    \quad
    -\left(r_{\uo\ut}(z,z^\prime)+r_{\ut\uo}(z^\prime, z)\right)\partial_{x}\delta(x{-}x')\, ,
\end{align}
where $\uo$ and $\ut$ refer to the two different factors of the tensor product $\mathfrak{g}\otimes \mathfrak{g}$ and $r_{\uo\ut}(z,z')$ is the classical $r$-matrix of the integrable system. 
In general, the proper bracket to use in the analysis is the Dirac bracket on reduced phase space. Lax integrability together with the Maillet form of Hamiltonian integrability are sufficient to prove the existence of an infinity of conserved charges that are in involution~\cite{Maillet:1985ec,Maillet:1986}.

We use the canonical formalism of~\cite{Matschull:1994vi} for coset models. 
This means that we introduce spurious coordinates $u^r$ with $r=1,\ldots, \dim(\mf{k})$ for the directions along $K$, while the physical fields will be called $\varphi^m$ with $m=1,\ldots, \dim(\mf{p})$. The collection of all fields is $\Phi^M=(\varphi^m, u^r)$ and we introduce a triangular vielbein on $G$\footnote{This is nothing but $\mathcal{V}$ written in the adjoint representation and with unfixed $K$-directions. }
\begin{align}
E_M{}^A=\begin{pmatrix}E_m{}^a(\varphi, u)&E_m{}^{\dot{a}}(\varphi, u)\\ 0& E_r{}^{\dot{a}}(u)\end{pmatrix}
\end{align}
that is also labelled by a flat index $A$ with a similar split $A=(a,\dot{a})$.
The inverse vielbein is written as
\begin{align}
E_A{}^M=\begin{pmatrix}E_a{}^m(\varphi, u)&\tilde{E}_a{}^r(\varphi, u)\\ 0& E_{\dot{a}}{}^r(u)\end{pmatrix}\, ,
\end{align}
with $\tilde{E}_a{}^r=-E_a{}^mE_{\dot{a}}{}^rE_m{}^{\dot{a}}$. 
In the vielbein formalism the structure constants of the algebra $[T_A, T_B]=f_{AB}{}^C T_C$ are related to the coefficients of anholonomy 
\begin{align}
\label{eq:anhol}
    \Omega_{AB}{}^C = 2 E_{[A}{}^M E_{B]}{}^N \partial_M E_N{}^C
\end{align}
by $f_{AB}{}^C = - \Omega_{AB}{}^C$. Note that the $\mathbb{Z}_2$-grading of the symmetric space forces certain structure constants to vanish (see appendix \ref{sec:appA}).

Using the notation $P_\alpha+Q_\alpha = P_\alpha^a\, T_a + Q_\alpha^{\dot{a}}\, T_{\dot{a}}$, where $T^a$ span $\mathfrak{p}$ and $T^{\dot{a}}$ span $\mathfrak{k}$, 
we obtain the components
\begin{align}
\label{eq:PE}
P^a_\alpha&=E_m{}^a(\varphi, u) \partial_\alpha \varphi^m\, ,\\
Q^{\dot{a}}_\alpha&=E_m{}^{\dot{a}}(\varphi, u)\partial_\alpha\varphi^m+ E_r{}^{\dot{a}}(u) \partial_\alpha u^r\, .
\end{align}

With the vielbein parametrisation~\eqref{eq:PE}, the position variables appearing in~\eqref{eq:undeformedLagrangianP} are $(\varphi^m, u^r)$. Instead of using their canonical conjugates, we prefer to use the following momentum-type variables
\begin{align}
\label{eq:pivar}
\pi_a&=E_a{}^m \frac{\partial\mathcal{L}}{\partial \dot{\varphi}^m}=  \gamma_{ab} P_{t}^b\, ,\\
\pi_{\dot{a}} &= E_{\dot{a}}{}^r \frac{\partial \mathcal{L}}{\partial \dot{u}^r} = 0 \,,
\end{align}
where $\gamma_{ab}$ is the invariant metric underlying $\Tr$.
The variable $\pi_{\dot{a}}$ is a first-class constraint that is associated with the local $K$-invariance of the system. 
The non-vanishing Poisson brackets of these variables are  
\begin{subequations}
\begin{align}
\lb \varphi^m(x), \pi_a(y)\rb_\PB &= E_a{}^m(x) \delta(x{-}y)\,,\\
\lb u^r(x), \pi_{\dot{a}}(y)\rb_\PB &= E_{\dot{a}}{}^r(x) \delta(x{-}y)\,,\\
\lb \pi_a(x), \pi_b(y)\rb_\PB &=  - f_{ab}{}^{\dot{c}} \pi_{\dot{c}}(x) \delta(x{-}y)\,,\\
\lb \pi_a(x), \pi_{\dot{b}}(y)\rb_\PB &=  - f_{a\dot{b}}{}^{c} \pi_{c}(x) \delta(x{-}y)\,,\\
\lb \pi_{\dot{a}}(x), \pi_{\dot{b}}(y)\rb_\PB &=  - f_{\dot{a}\dot{b}}{}^{\dot{c}} \pi_{\dot{c}}(x) \delta(x{-}y)\,,
\end{align}
\end{subequations}
where we have used the structure constants in the symmetric space split $A=(a, \dot{a})$ as well as~\eqref{eq:anhol}. 

The above relations lead to the equal time Poisson brackets
\begin{subequations}
\label{eq:Poiss1}
\begin{align}
\lb Q_x^{\dot{a}}(x), Q_x^{\dot{b}}(y)\rb_{\PB} &=\lb P_x^a(x), P_x^b(y)\rb_\PB =0\, ,\\
\lb P_t^a(x), P_t^b(y)\rb_\PB &=
-f^{ab}{}_{\dot{c}}\gamma^{\dot{c}\dot{d}}\pi_{\dot{d}}(x)\delta(x{-}y)\, ,\\
\lb P_t^a(x),P_x^b(y)\rb_\PB &=
-f^{ab}{}_{\dot{c}}Q_x^{\dot{c}}(x)\delta(x{-}y)+\gamma^{ab}\partial_x\delta(x{-}y)\, ,\\
\lb P_t^a(x), Q_x^{\dot{b}}(y)\rb_\PB &= -f^{a\dot{b}}{}_{c} P_x^c(x) \delta(x{-}y)\, ,\\
\gamma^{\dot{c}\dot{d}}\lb \pi_{\dot{c}}(x),Q_x^{\dot{a}}(y)\rb_\PB &=-f^{\dot{d}\dot{a}}{}_{\dot{b}}Q_x^{\dot{b}}(x)\delta(x{-}y)+\gamma^{\dot{d}\dot{a}}\partial_x\delta(x{-}y)\, ,\\
\gamma^{\dot{c}\dot{d}}\lb P_t^a(x),\pi_{\dot{c}}(y)\rb_\PB &= -f^{a\dot{d}}{}_d P_t^d(x)\delta(x{-}y)\, ,\\
\gamma^{\dot{c}\dot{d}}\lb\pi_{\dot{c}}(x),P_x^{a}(y)\rb_\PB &=-f^{\dot{d}a}{}_b P_x^b(x) \delta (x{-}y)\, ,\\
\gamma^{\dot{c}\dot{d}}\gamma^{\dot{a}\dot{b}}\lb \pi_{\dot{c}}(x),\pi_{\dot{a}}(y)\rb_\PB &=-f^{\dot{d}\dot{b}}{}_{\dot{e}}\gamma^{\dot{e}\dot{f}}\pi_{\dot{f}}(x)\delta(x{-}y)\, .
\end{align}
\end{subequations}

In the Hamiltonian formalism, the spatial part of the Lax connection \eqref{eq:cosetLaxundeformed} for the coset sigma model \eqref{eq:undeformedLagrangianP} can be supplemented with a term proportional to the constraint $\pi_{\dot{a}}$, as this has no effect on the constraint submanifold of phase space \cite{Lacroix:2018njs},
\begin{align}
\label{eq:improvedLax}
\mathfrak{L}_x &=Q_x +\frac{1+z^2}{1-z^2}P_x+\frac{2z}{1-z^2} P_t + f(z) \tilde{\pi}\, ,
\end{align}
with $\tilde{\pi}=\gamma^{\dot{a}\dot{b}}\pi_{\dot{a}}T_{\dot{b}}$ and $f(z)$ to be determined. Its role is to arrive at an expression where the $r$-matrix is expressed in terms of the universal solution with a twist function.

Using~\eqref{eq:improvedLax} together with the Poisson brackets~\eqref{eq:Poiss1} one can show that the Maillet form~\eqref{eq:Maillet} is satisfied with the $r$-matrix \cite{Sevostyanov:1995hd,Delduc:2012qb} \begin{align}
\label{eq:r0}
    r_{\uo\ut}(z,z^\prime)=\varphi^{-1}(z^\prime)\left(g(z, z^\prime)\CCo-g(z^\prime, z)\CCt\right)\, ,
\end{align}
where
\be\label{eq:g0}
g(z, z^\prime)= \frac{\left(z^2-1\right) (z^\prime-1)^2}{4 (z-z^\prime) (z z^\prime-1)}\, ,
\ee
is part of the universal solution that is augmented by the `twist' function 
\be\label{eq:twist0}
\varphi(z^\prime)=-\frac{(z^\prime-1)^2 ({z^\prime}^2-1)}{16 {z^\prime}^2}\,.
\ee
The free function $f(z)$ appearing in~\eqref{eq:improvedLax} has to be chosen as
\be\label{eq:f0}
f(z)=\frac{2 z}{(z-1)^2}\, .
\ee
The (split) Casimir appearing in~\eqref{eq:r0} is defined from
\begin{align}
C_{\underline{\mathbf{12}}}=\gamma^{AB}T_A\otimes T_B=\gamma^{ab}T_a\otimes T_b+\gamma^{\dot{a}\dot{b}}T_{\dot{a}}\otimes T_{\dot{b}}=\overset{\scriptscriptstyle (1)}{C}_{\underline{\mathbf{12}}}+\overset{\scriptscriptstyle (0)}{C}_{\underline{\mathbf{12}}}
\end{align}
and satisfies 
\begin{align}
&[X_{\underline{\mathbf{1}}}, \overset{\scriptscriptstyle{(1)}}C_{\underline{\mathbf{1}}\underline{\mathbf{2}}}]=-[X_{\underline{\mathbf{2}}}, \overset{\scriptscriptstyle{(0)}}C_{\underline{\mathbf{1}}\underline{\mathbf{2}}}]\, ,\quad \quad [\tilde{X}_{\underline{\mathbf{1}}}, \overset{\scriptscriptstyle{(1)}}C_{\underline{\mathbf{1}}\underline{\mathbf{2}}}]=-[\tilde{X}_{\underline{\mathbf{2}}}, \overset{\scriptscriptstyle{(1)}}C_{\underline{\mathbf{1}}\underline{\mathbf{2}}}]\, ,\\
&[X_{\underline{\mathbf{1}}}, \overset{\scriptscriptstyle{(0)}}C_{\underline{\mathbf{1}}\underline{\mathbf{2}}}]=-[X_{\underline{\mathbf{2}}}, \overset{\scriptscriptstyle{(1)}}C_{\underline{\mathbf{1}}\underline{\mathbf{2}}}]\, ,\quad \quad [\tilde{X}_{\underline{\mathbf{1}}}, \overset{\scriptscriptstyle{(0)}}C_{\underline{\mathbf{1}}\underline{\mathbf{2}}}]=-[\tilde{X}_{\underline{\mathbf{2}}}, \overset{\scriptscriptstyle{(0)}}C_{\underline{\mathbf{1}}\underline{\mathbf{2}}}]\,, 
\end{align}
for two objects $X=X^aT_a$ and $\tilde{X}=\tilde{X}^{\dot{a}}T_{\dot{a}}$, 
which are important for proving the Maillet form. An important aspect of the proof is that the term $P_t$ in the Lax connection~\eqref{eq:improvedLax} is basically the momentum of the physical fields according to~\eqref{eq:pivar}. 
In the next section, we will consider an integrable deformation of the system~\eqref{eq:undeformedLagrangianP} and integrability will rely on the fact that this property of the Lax connection is preserved.

\section{Deformations}
\label{sec:Z2def}

In this section, we consider a general family of deformations of the coset model, similar in spirit to the principal chiral model analysis of~\cite{Ferko:2024ali}. We first present the Lorentz-covariant deformation at the Lagrangian level and exhibit an associated Lax connection that reproduces the equations of motion. Then we prove Hamiltonian integrability in Maillet form through a canonical analysis. 

\subsection{Lagrangian deformation}

We consider the following deformation of the Lagrangian~\eqref{eq:undeformedLagrangianP} 
\be\label{eq:deformedLagrangianv}
\widetilde{\mathcal{L}}=\frac{1}{2}\text{Tr}(P_\alpha P^\alpha) + \text{Tr}(2P_\alpha v^\alpha +v_\alpha v^\alpha)+E(\nu)\, .
\ee
Here, $v_\alpha \in \mathfrak{p}$ is an auxiliary vector that only appears algebraically in the Lagrangian and the deformation is parametrised by an arbitrary scalar function  $E(\nu)$  of the deformation $\nu$ that is the defined as the  scalar combination
\begin{align}
\label{eq:nudef}
\nu=(2\eta^{\alpha\beta}\eta^{\gamma\delta}-\eta^{\alpha\gamma}\eta^{\beta\delta} )\text{Tr}(v_\alpha v_\gamma)\text{Tr}(v_\beta v_\delta)=\text{Tr}(v_+v_+)\text{Tr}(v_-v_-)\,.
\end{align}
In the last equality we have used light-cone components defined with respect to the light-cone coordinates
\begin{align}
x^\pm = \frac12 (t\pm x)\,.
\end{align}
The choice of deformation parameter~\eqref{eq:nudef} follows~\cite{Ferko:2024ali} where it is motivated from modified Maxwell theories. In section~\ref{sec:Zn}, we will argue from a different perspective why the above choice is natural and from an integrable systems point of view, combined with manifest Lorentz invariance.

The equations of motion of the model \eqref{eq:deformedLagrangianv} are now
\begin{subequations}
\label{eq:eom2}    
\begin{align}
\label{eq:algebraicauxdef}
P_\alpha  &= - v_\alpha - 2 E'(\nu)  (2\eta^{\beta\gamma} \delta_\alpha^\delta -\eta^{\beta\delta} \delta^\gamma_\alpha ) \Tr(v_\beta v_\delta) v_\gamma\,,\\
\label{eq:dynamicauxdef}
D_\alpha\mathcal{P}^\alpha &=0\, ,
\end{align}
\end{subequations}
where we have defined 
\begin{align}
\label{eq:Jdef}
\mathcal{P}_\alpha=-(P_\alpha+2 v_\alpha)\, .
\end{align}

The parameter $v_\alpha$  appears algebraically in the Lagrangian and can in principle be integrated out by substituting the solution to its equation of motion into the Lagrangian. However, this is generally not possible to do explicitly. If there is no deformation ($E\equiv 0$), the solution is $P_\alpha = -v_\alpha$ and~\eqref{eq:deformedLagrangianv} becomes the orginal coset Lagrangian~\eqref{eq:undeformedLagrangianP}. Moreover, the quantity $\mathcal{P}_\alpha=P_\alpha$ in this case. As we will see below, the temporal component $\mathcal{P}_t$ is a suitable substitute for $P_t$ in the canonical formalism in the general case.

Even with a general deformation, the system admits a Lax description.
At the Lorentz-covariant level, we can consider the Lax connection 
\begin{equation}
\label{eq:newLaxdefaux}
\widetilde{\mathfrak{L}}_\alpha(z)\equiv Q_\alpha +\frac{1+z^2}{1-z^2} P_\alpha +\frac{2z}{1-z^2} \epsilon_{\alpha\beta}\eta^{\beta\gamma}\mathcal{P}_\gamma,
\end{equation}
whose compatibility condition 
\be
2\partial_{[\alpha}\widetilde{\mathfrak{L}}_{\beta]}+[\widetilde{\mathfrak{L}}_\alpha, \widetilde{\mathfrak{L}}_\beta]=0\, 
\ee
implies both the dynamical equation of motion \eqref{eq:dynamicauxdef} as well as the algebraic equation of motion \eqref{eq:algebraicauxdef} as well as the Bianchi identitites \eqref{eq:Bianchis}. In particular, there is no need to separate the two equations of motion~\eqref{eq:eom2} in the analysis.

\subsection{Hamiltonian integrability}

The set of canonical variables of~\eqref{eq:deformedLagrangianv} now also includes $v_\alpha\in\mathfrak{p}$ and its conjugate momentum $\mathsf{p}^\alpha$. Since $v_\alpha$ appears only algebraically in the Lagrangian, one has the constraint that its conjugate momentum vanishes: $\mathsf{p}^\alpha = 0$.

We again use the following momentum-type variable associated with the physical scalars
\begin{align}
\label{eq:calPdef}
\pi_a=E_a{}^m p_m=E_a{}^m \frac{\partial\widetilde{\mathcal{L}}}{\partial \dot{\varphi}^m}=\gamma_{ab}\mathcal{P}_{\tau}^b\, .
\end{align}
Comparing to~\eqref{eq:pivar}, we see that  $\mathcal{P}_t^a$ takes the place of  $P_t^a$ in the canonical formalism.
Then one can check that the Poisson brackets take the same form as in~\eqref{eq:Poiss1} except for the change that $P_t^a$ is replaced by $\mathcal{P}_t^a$ consistently:
\begin{subequations}
\label{eq:Poiss2}
\begin{align}
\lb Q_x^{\dot{a}}(x), Q_x^{\dot{b}}(y)\rb_\PB &=\lb P_x^a(x), P_x^b(y)\rb_\PB =0\, ,\\
\lb \mathcal{P}_t^a(x), \mathcal{P}_t^b(y)\rb_\PB &=
-f^{ab}{}_{\dot{c}}\gamma^{\dot{c}\dot{d}}\pi_{\dot{d}}(x) \delta(x{-}y)\, ,\\
\lb \mathcal{P}_t^a(x),P_x^b(y)\rb_\PB &=-
f^{ab}{}_{\dot{c}}Q_\sigma^{\dot{c}}(x) \delta(x{-}y)+\gamma^{ab}\partial_x\delta(x{-}y)\, ,\\
\lb \mathcal{P}_t^a(x), Q_x^{\dot{b}}(y)\rb_\PB &=-f^{a\dot{b}}{}_{c} P_\sigma^c(x) \delta(x{-}y)\, ,\\
\gamma^{\dot{c}\dot{d}}\lb \pi_{\dot{c}}(x),Q_\sigma^{\dot{a}}(y)\rb_\PB &=-f^{\dot{d}\dot{a}}{}_{\dot{b}}Q_\sigma^{\dot{b}}(x)\delta(x{-}y)+\gamma^{\dot{d}\dot{a}}\partial_x\delta(x{-}y)\, ,\\
\gamma^{\dot{c}\dot{d}}\lb \mathcal{P}_t^a(x),\pi_{\dot{c}}(y)\rb_\PB &= -f^{a\dot{d}}{}_d\mathcal{P}_t^d(x) \delta(x{-}y)\, ,\\
\gamma^{\dot{c}\dot{d}}\lb \pi_{\dot{c}}(x),P_x^{a}(y)\rb_\PB &=-f^{\dot{d}a}{}_b P_x^b(x) \delta(x{-}y)\, ,\\
\gamma^{\dot{c}\dot{d}}\gamma^{\dot{a}\dot{b}}\lb \pi_{\dot{c}}(x),\pi_{\dot{a}}(y)\rb_\PB &=-f^{\dot{d}\dot{b}}{}_{\dot{e}}\gamma^{\dot{e}\dot{f}} \pi_{\dot{f}}(x) \delta(x{-}y)\, .
\end{align}
\end{subequations}

We next discuss the structure of the constraints of the theory.
There is still the first-class constraint $\pi_{\dot{a}}=0$ that is associated with the local $K$-invariance. 

The conservation of the new primary constraint $\mathsf{p}^\alpha=0$ triggers a secondary constraint that is equivalent to the algebraic equation of motion~\eqref{eq:algebraicauxdef}. Due to the appearance of $v_\alpha$ in~\eqref{eq:algebraicauxdef}, this secondary constraint does not Poisson commute with $\mathsf{p}^\alpha$ and these two constraints together form a pair of second-class constraints. 
The secondary constraint~\eqref{eq:algebraicauxdef} does not Poisson commute with itself either, so that we have the schematic form
\begin{align}
    \lb \psi_i, \psi_j \rb = M_{ij} = \begin{pmatrix}
        0 & *\\
        * & *
    \end{pmatrix}\, ,
\end{align}
with $\psi_1=\mathsf{p}^\alpha$ and $\psi_2$ corresponding to~\eqref{eq:algebraicauxdef}. The matrix $M_{ij}$ needs to be inverted to obtain the Dirac brackets of the theory. This is a non-trivial task which is fortunately not needed for proving Hamiltonian integrability. We shall see that we only need the fact that
\begin{align}
(M^{-1})^{ij} =\begin{pmatrix}
    * & * \\
    * &0
\end{pmatrix}    \,,
\end{align}
so that at least one of the constraints contracting $M^{-1}$ has to be $\psi_1=\mathsf{p}^\alpha$.

The modified Lax connection in Hamiltonian form is now (with $\mathcal{P}_t = \mathcal{P}_t^a T_a$)
\begin{align}
\label{eq:defLaxG/HHamiltonian}
\widetilde{\mathfrak{L}}_x &=Q_x +\frac{1+z^2}{1-z^2}P_x+\frac{2z}{1-z^2}\mathcal{P}_t + f(z) \tilde{\pi}\, .
\end{align}
Using the Poisson brackets~\eqref{eq:Poiss2} one can show as before that this modified Lax connection satisfies the Maillet integrability condition for exactly the same $r$-matrix and twist function as in the undeformed theory. The bracket required for the Maillet form in this case is now the Dirac bracket in view of the second-class constraints associated with the auxiliary field $v_\alpha$.
However, when computing the Dirac brackets of $\widetilde{\mathfrak{L}}_x$ with itself, all the terms going beyond the standard Poisson bracket are constrained by the structure of $M^{-1}$ and contain a Poisson commutator with $\mathsf{p}^\alpha$ that vanishes.\footnote{We thank Christian Ferko and Liam Smith for clarifying this point.}
Here, it is important to replace any occurrence of $\mathcal{P}_t^a$ by $\gamma^{ab} E_a{}^m p_m$ where $p_m$ is the canonical momentum of the physical scalars according to~\eqref{eq:calPdef}, so that $\lb \mathcal{P}_t^a, \mathsf{p}^\alpha\rb =0$.
For this reason, the proof of Hamiltonian integrability can be based on the Poisson brackets~\eqref{eq:Poiss2} and follows the proof of the undeformed model. This argument is analogous to the one in the principal chiral model studied in~\cite{Ferko:2024ali}.

In summary, the method of~\cite{Ferko:2024ali} is also applicable to symmetric space coset models and generates an infinity of classically integrable models. In the next section we will extend this to $\mathbb{Z}_N$-coset spaces and use a slightly different strategy that does, a priori, not rely on the Lagrangian deformation $E(\nu)$.


\section{\texorpdfstring{Generalisation to $\mathbb{Z}_N$-coset models}{Generalisation to Z(n)-coset models}} 
\label{sec:Zn}
A natural question is whether the previous result can be generalised from (Riemannian) symmetric spaces to (semi-symmetric) $\mathbb{Z}_N$-cosets. Indeed this is possible. To this end we will also demonstrate here an alternative logic, than the one in \cite{Ferko:2024ali}, that is applicable to the principle chiral model but also to the $\mathbb{Z}_N$-coset model. I.e. we will use the fact that the auxiliary field deformations of both the principal chiral model \cite{Ferko:2024ali} and the symmetric homogeneous space (section \ref{sec:Z2def}) take a common form inside the Lax formalism. Our procedure will consist of the following steps:
\begin{enumerate}
    \item Deformations of a known Lax connection (schematically for principal chiral model and $\mathbb{Z}_N$-coset $\sigma$-models) 
        \begin{equation}
            \mathfrak{L}(z) = Q + \alpha(z) \ P + \beta(z)  \star P \quad \rightarrow \quad \widetilde{\mathfrak{L}}(z) = Q + \alpha(z) \ P + \beta(z)  \star \mathcal{P}\,,
        \end{equation}
        (where $Q=0$ in the principal chiral model) and 
        with an, a priori, independent current~$\mathcal{P}$.
    \item Flatness of the Lax connection will be equivalent to equations of motion for $\mathcal{P}$, the Maurer--Cartan equation for $P$ and some algebraic equations of motion (constraints) relating $P$ and $\mathcal{P}$ (schematically)
    \begin{equation}
        [\mathcal{P},\mathcal{P}] = [P,P]\,, \qquad [P,\star \mathcal{P}] =0\,.
    \end{equation}
    These constraints will have slightly more general solutions than the auxiliary field deformation in \cite{Ferko:2024ali}.

    \item Hamiltonian integrability is ensured by the fact the $\mathcal{P}_\tau = \pi$, with $\pi$ being related to the canonical momentum as in the undeformed case.

    \item Finding a Lagrangian, that reproduces the equations of motions and $\mathcal{P}_\tau$ as canonical momentum, is then not automatic. In principle, this could be achieved by hand, as the Hamiltonian can be computed as one of the conserved charges from the Lax formalism. We will present a Lagrangian realisation as natural generalisation of the auxiliary field deformation of the principal chiral model in \cite{Ferko:2024ali} to the $\mathbb{Z}_N$-coset $\sigma$-model.
\end{enumerate}

\subsection{Review of the undeformed case}

\paragraph{$\mathbb{Z}_N$-symmetric homogeneous spaces.} Again consider a $\sigma$-model with target space $M = G/K$, where $G$ and $K$ are Lie (super)groups with Lie (super)algebras $\mathfrak{g}$ resp. $\mathfrak{k} = \mathfrak{g}^{(0)}$ that are subject to a $\mathbb{Z}_N$-grading
\begin{align}
\mathfrak{g}  = \mathfrak{k} \oplus \mathfrak{p} &= \mathfrak{g}^{(0)} \oplus \mathfrak{g}^{(1)} \oplus ... \oplus \mathfrak{g}^{(N-1)}\, , 
\end{align}
with $[\mathfrak{g}^{(i)},\mathfrak{g}^{(j)}] \subset \mathfrak{g}^{(i+j \ \text{mod} N)}$ and $\mathfrak{p} = \bigoplus_{i>0} \mathfrak{g}^{(i)}$. This can be formalised by a Lie algebra automorphism $\sigma: \ \mathfrak{g}\rightarrow \mathfrak{g}$ with $\sigma^N = 1$ and the $\mathfrak{g}^{(i)}$ being the eigenspaces of that automorphism. For a generic Lie algebra $\mathfrak{g}$ and $N >2$
\begin{equation}
[\mathfrak{k} , \mathfrak{k}] \subset \mathfrak{k}, \quad [\mathfrak{k},\mathfrak{p}] \subset \mathfrak{p} \quad \text{and} \quad [\mathfrak{p},\mathfrak{p}] \not\subset \mathfrak{k}.
\end{equation}
The latter makes the difference to the previous case of a symmetric homogeneous space (i.e. a $\mathbb{Z}_2$-coset). Several descriptions of such spaces exist in the literature, like \textit{non-symmetric} or \textit{semi-symmetric} homogeneous space. For brevity, these spaces are referred to as \textit{$\mathbb{Z}_N$-cosets} in this paper. 

As usual, in order to construct a $G$-invariant action quadratic in $\mathfrak{g}$-valued currents, the existence of a non-degenerate Ad-invariant bilinear form on $\mathfrak{g}$ is required. This has to respect its $\mathbb{Z}_N$-grading: 
\begin{equation}
\text{Tr}\left( m n \right) = \sum_{i = 0}^{N-1}\text{Tr}\left( m^{(i)} n^{(N-i)} \right)\, , \quad \text{with} \quad m^{(i)},n^{(i)} \in \mathfrak{g}^{(i)}.
\end{equation}
Here it is denoted by `Tr', but it can be any bilinear form with these properties. All calculations here hold for Lie superalgebras with a $\mathbb{Z}_{2N}$-grading as well, where it would be the supertrace.

\paragraph{The action.} The crucial difference with the principal chiral model or the symmetric ($\mathbb{Z}_2$-) coset, is that it is possible to introduce a certain $b$-field term to the action. This will also turn out to be necessary in order to achieve classical integrability. In correspondence to the symmetric space case, we expand the Maurer--Cartan form of $G$-valued fields $\mathcal{V}$ into $\mathbb{Z}_N$-eigenspaces
\begin{equation}
\mathcal{V}^{-1} \mathrm{d} \mathcal{V} = Q + \sum_{i=1}^{N-1} P^{(i)}, \qquad Q \in \mathfrak{k}=\mathfrak{g}^{(0)}, \ P^{(i)} \in \mathfrak{g}^{(i)} .
\end{equation}
The Maurer--Cartan (Bianchi) identities of these current components take the following form:
\begin{subequations} \label{eq:MaurerCartanZN}
\begin{align}
    0 &= \mathrm{d} Q + \frac{1}{2} [Q,Q] + \frac{1}{2} \sum_{i=1}^{N-1} [P^{(i)},P^{(N-i)}]\, , \label{eq:MaurerCartanZN1} \\
    0 &= \mathrm{D} P^{(i)} + \frac{1}{2} \sum_{j\neq i} [P^{(j)},P^{(i-j)}]\, , 
\end{align}
\end{subequations}
where we introduced the $K$-covariant derivative $D = \mathrm{d} + [Q , \ ]$.

A general action, quadratic in currents and invariant under the usual global $G$- and local $K$-transformations \eqref{eq:costrm}, looks as follows: 
\begin{equation}
\mathcal{L} = \frac{1}{2} \text{Tr} \left(P_+ \pi_- (P_-) \right), \label{eq:ZNLagrangian}
\end{equation}
with 
\begin{equation}
    \pi_\pm = \sum_{i=1}^{N-1} (s_i \pm b_i) \pi^{(i)}
\end{equation}  
and $\text{Tr}(\pi_+(m) n) = \text{Tr}(m \pi_-(n) )$, where the constants $s_i = s_{N-i}$ and $b_i = - b_{N-i}$, for $i = 1,...,N-1$ are factors in front of kinetic resp. Wess--Zumino-terms, and $\pi^{(i)}$ are projectors onto $\mathfrak{g}^{(i)}$. I.e. the Lagrangian can be written as
\begin{equation}
\mathcal{L} \sim - \sum_i \text{Tr}\left( s_i P^{(i)} \wedge \star P^{(N-i)} + b_i P^{(i)} \wedge P^{(N-i)} \right).
\end{equation}
For $N = 2$, the case of sections \ref{sec:rev} and \ref{sec:Z2def}, there is no admissible $b$-field of this type. This complicates the analysis for $N>2$ slightly.

The equations of motion are (for $i=1,...,N-1$):
\begin{equation}
    0 = D\star K^{(i)} + [P,K]^{(i)}, \qquad \text{with} \quad K^{(i)} = s_i P^{(i)} + b_i \star P^{(i)}.
\end{equation}

\paragraph{Integrable choices for $s_i$ \& $b_i$.} Different choices of the parameters $s_i$ and $b_i$ correspond to distinct models and, in contrast to the symmetric space $\sigma$-model, not all choices correspond to an integrable model. The two main classes that have been studied in the literature \cite{Young:2005jv,Kagan:2005wt,Ke:2008zz,Ke:2011zzb,Ke:2011zzc,Ke:2011zzd,Bykov:2014efa,Bykov2016cyclic,Bykov:2016rdv,Ke:2017wis,Bykov:2018vbg,Delduc:2019lpe,Hoare:2021dix} are the `pure spinor'- and `Green--Schwarz'-type, that are named such as generalisation of the form of the $\mathbb{Z}_4$-supercoset $\sigma$-model. Other instances of integrable choices of $s_i$ and $b_i$ have been studied in some but not full generality in \cite{Osten:2021opf}.

For simplicity, here we will only consider the so-called `pure spinor'-type choice \cite{Young:2005jv}:
\begin{equation}
    s_i \equiv 1, \qquad b_i = \left( 1 - \frac{2i}{N} \right).
\end{equation}
This model possesses the Lax representation with spectral parameter $\lambda$
\begin{equation}
    \mathfrak{L}(\lambda) = Q + \frac{1}{2} \sum_{i=1}^{N-1} \frac{\lambda^N + 1}{\lambda^{N-i}} P^{(i)} + \frac{1}{2} \sum_{i=1}^{N-1} \frac{\lambda^N - 1}{\lambda^{N-i}} \star P^{(i)}\, ,
\end{equation}
and $r$-matrix \cite{Ke:2011zzb,Lacroix:2018njs}
\begin{equation}\label{eq:rmatixdeformedZN}
    r_{\uo \ut}(\lambda,\mu) = \varphi^{-1}(\mu) \sum_{i=0}^{N-1} \frac{\lambda^i \mu^{N-i-1} \overset{(i)}{C}_{\uo \ut}}{\lambda^N - \mu^N}\,,
\end{equation}
for a Maillet bracket \eqref{eq:Maillet}. In \eqref{eq:rmatixdeformedZN}, $\overset{(i)}{C}_{\uo \ut} = \pi^{(i)} {C}_{\uo \ut}$ and the twist function has the form
\begin{equation}\label{eq:twistZN}
    \varphi(\lambda) = - \frac{2 \lambda^{N-1}}{(\lambda^N - 1)^2}\,.
\end{equation}
Let us note that, in comparison to the $\mathbb{Z}_2$-case \eqref{eq:cosetLaxundeformed}, we made the following reparameterisation of the spectral parameter:
\begin{equation}
    \lambda = \frac{1 + z}{1 - z} \,.
\end{equation}

\subsection{Integrability of the auxiliary field deformation}

Assume the generalisation of the Lax connection of the (pure spinor type) $\mathbb{Z}_N$-coset model (same conventions as in \cite{Osten:2021opf}):
\begin{equation}
    \widetilde{\mathfrak{L}}(\lambda) = Q + \frac{1}{2} \sum_{i=1}^{N-1} \frac{\lambda^N + 1}{\lambda^{N-i}} P^{(i)} + \frac{1}{2} \sum_{i=1}^{N-1} \frac{\lambda^N - 1}{\lambda^{N-i}} \star \mathcal{P}^{(i)}\,.
\end{equation}
Then, flatness of the Lax connection implies:
\begin{align}
    0 &= \mathrm{d}\widetilde{\mathfrak{L}}(\lambda) + \frac{1}{2} [\widetilde{\mathfrak{L}}(\lambda),\widetilde{\mathfrak{L}}(\lambda)] \displaybreak[2] \nn \\
    &= \mathrm{d} Q + \frac{1}{2} [Q , Q ] + \frac{1}{4} \sum_{j=1}^{N-1} \left( [P^{(j)} , P^{(N-j)}] + [\mathcal{P}^{(j)},\mathcal{P}^{(i-j)}] \right) \nn \\
    &{} \qquad + \frac{\lambda^{2N} + 1}{8 \lambda^N} \sum_j \left( [P^{(j)} , P^{(N-j)}] - [\mathcal{P}^{(j)},\mathcal{P}^{(N-j)}] \right) \nn\\
    &{} \qquad + \frac{1}{2} \sum_{i=1}^{N-1} \frac{\lambda^N + 1}{\lambda^{N-i}} \left( D P^{(i)} + \frac{1}{2} \sum_{j \neq 0,i} [P^{(j)} , P^{(i-j)}] \right) \displaybreak[2] \nn\\
    &{} \qquad + \frac{1}{2} \sum_{i=1}^{N-1} \frac{\lambda^N - 1}{\lambda^{N-i}} \left( D \star \mathcal{P}^{(i)} + \frac{1}{4} \left( - \sum_{j<i} + \sum_{j>i} \right)  \left( [P^{(j)} , P^{(i-j)}] + [\mathcal{P}^{(j)},\mathcal{P}^{(i-j)}] \right) \right) \nn\\
    &{} \qquad - \frac{1}{8} \sum_{i=1}^{N-1} \frac{\lambda^N - 1}{\lambda^{N-i} } \left(\frac{1}{\lambda^N} \sum_{j<i} + \lambda^N \sum_{j>i} \right) \left( [P^{(j)} , P^{(i-j)}] - [\mathcal{P}^{(j)},\mathcal{P}^{(i-j)}] \right) \nn \\ 
    &{} \qquad + \frac{1}{4} \sum_{i,j >0 } \frac{(\lambda^N -1)(\lambda^N + 1)}{\lambda^{N-i} \lambda^{N-j}} [P^{(i)}, \star \mathcal{P}^{(j)}] \,.
\end{align}
Flatness for all $\lambda$ implies the following constraints (algebraic equations of motion):
\begin{align}
    0 &= \sum_{j=1}^{N-1} \left( [P^{(j)} , P^{(N-j)}] - [\mathcal{P}^{(j)},\mathcal{P}^{(N-j)}] \right)\,, \nonumber \\
    0 &= \sum_{j<i} \left( [P^{(j)} , P^{(i-j)}] - [\mathcal{P}^{(j)},\mathcal{P}^{(i-j)}] \right) = \sum_{i > j}  \left( [P^{(j)} , P^{(i-j)}] - [\mathcal{P}^{(j)},\mathcal{P}^{(i-j)}] \right)\,,  \label{eq:Constraints} \\
    0 &= [P^{(i)}, \star \mathcal{P}^{(j)}] + [P^{(j)}, \star \mathcal{P}^{(i)}] \,.\nonumber
\end{align}
With these constraints the remaining equations of motions deduced from the flatness of the Lax connection become the Maurer--Cartan identity \eqref{eq:MaurerCartanZN} and what will become the equations of motion
\begin{align}
    0 &= D \star \mathcal{P}^{(i)} + \frac{1}{2} \left( - \sum_{j<i} + \sum_{j>i} \right)  [P^{(j)} , P^{(i-j)}]  \,.\label{eq:ZNdynamicaleom}
\end{align}
In principle any solution to the constraints \eqref{eq:Constraints} should give a well-defined integrable model. A general solution (slightly more general than the $E$-deformation) is
\begin{equation}
    \mathcal{P} = \pm ( P + 2v ), \qquad P_\pm^{(i)} = - v_\pm^{(i)} + f^{(\pm,i)}(v_+ , v_-) v_\mp^{(i)}\, , \label{eq:generalDef}
\end{equation}
with $f^{(\pm)} \equiv 0$ corresponding to the undeformed $\mathbb{Z}_N$-coset model. The crucial question is what general form such a deformation might have on the Lagrangian level. This question will be briefly addressed below in section \ref{sec:ZNLagRep}.

Different Lax connections for the $\mathbb{Z}_N$-coset $\sigma$-models of Green--Schwarz or hybrid type \cite{Osten:2021opf} would be expected to be valid starting points for deformations of the same kind. For simplicity of presentation they are not investigated here.

\paragraph{Hamiltonian integrability.} The question is whether the Poisson bracket of the Lax matrix 
\begin{align}
    \widetilde{\mathfrak{L}}_\sigma (\lambda) = Q_\sigma + \sum_{i=1}^{N-1} \left( \frac{\lambda^N + 1}{\lambda^{N-i}} P_\sigma^{(i)} + \frac{\lambda^N - 1}{\lambda^{N-i}} \mathcal{P}_\tau \right) + f(\lambda) \pi^{(0)}
\end{align}
can be put into Maillet form \eqref{eq:Maillet}. For this, again the last term can be added as the constraint associated to account for local $K$-symmetry, and bring the $r$-matrix into an appropriate form.

The crucial condition to quickly check Hamiltonian integrability is
\begin{equation}
    \mathcal{P}_\tau^{(i),a} = \pi^{(i),a}\, ,
\end{equation}
with the latter being the canonical momentum (twisted by the vielbein). This is an additional non-trivial condition that one has to impose in order to ensure integrability of the deformed model.

Then the Lax matrix of the auxiliary field deformed $\mathbb{Z}_N$-cosets can be written as
\begin{align}
    \widetilde{\mathfrak{L}}_\sigma(\lambda) = Q_\sigma + \sum_{i=1}^{N-1} \left( \frac{\lambda^N + 1}{\lambda^{N-i}} P_\sigma^{(i)} + \frac{\lambda^N - 1}{\lambda^{N-i}} \pi \right) + f(\lambda) \pi^{(0)} \, .
\end{align}
Hence, phrased in canonical variables there is no difference in comparison to the undeformed case. Also, all considerations regarding the constraint analysis and most crucially the fact that 
\begin{align*}
    \{P_\sigma^{(i)}(x) , P_\sigma^{(j)}(y) \}_\DB &= \{P_\sigma^{(i)}(x) , P_\sigma^{(j)}(y) \}_\PB, \\
    \{P_\sigma^{(i)}(x) , \pi^{(j)}(y) \}_\DB &= \{P_\sigma^{(i)}(x) , \pi^{(j)}(y) \}_\PB, \\
    \{ \pi^{(i)}(x) , \pi^{(j)}(y) \}_\DB &= \{ \pi^{(i)}(x) , \pi^{(j)}(y) \}_\PB ,
\end{align*}
go through from the considerations in \cite{Ferko:2024ali} and above in section \ref{sec:Z2def}. With this the proof of the involution of conserved charges via the Maillet bracket \eqref{eq:Maillet} and the computation of an $r$-matrix of twist function form, is the same as in the undeformed case, with the result given above, \eqref{eq:rmatixdeformedZN} and \eqref{eq:twistZN}. 

\subsection{Lagrangian realisation} \label{sec:ZNLagRep}
So far, we have constructed a general set of (Hamiltonian) integrable sets of equations of motion. Now, let us present a Lagrangian description of the above system that corresponds to the generalisation of the auxiliary field deformation of the principal chiral model in \cite{Ferko:2024ali}:
\begin{equation}
     \widetilde{\mathcal{L}} \sim - \frac{1}{2} \text{tr} \left( P_+ \pi_- P_- \right)  + \sum_{i=1}^{N-1} s_i \text{tr}\left( P_+^{(i)} v_-^{(N-i)} + v_+^{(N-i)} P_-^{(i)} + v_+^{(i)} v_-^{(N-i)} \right) + E(\xi^{(+)},\xi^{(-)}). \label{eq:ZNdefaction}
 \end{equation}
 The auxiliary fields $v = \sum_{i=1}^{N-1} v^{(i)}$ are valued in $\mathfrak{p}$ and we defined
 \begin{equation}
     \xi^{(\pm)} = \sum_{i=1}^{N-1} s_i \text{Tr}(v_\pm^{(i)} v_\pm^{(N-i)}).
 \end{equation}
The deformed Lagrangian of the $\mathbb{Z}_2$-case \eqref{eq:defLaxG/HHamiltonian} is reproduced for the case $E (\xi^{(+)} , \xi^{(-)}) = E (\xi^{(+)} \xi^{(-)})$.
 
Writing $K=\sum_{i=1}^{N-1} K^{(i)}$ and $P=\sum_{i=1}^{N-1} P^{(i)}$ ($P^{(0)} = K^{(0)} = 0$), the equations of motion are
 \begin{align}
     0 &= D \star K^{(i)} + [P, \star K]^{(i)}, \quad \text{with} \quad K^{(i)} = - s_i \mathcal{P}^{(i)} - b_i \star P^{(i)}, \quad K_\pm^{(i)} = \pi_\pm^{(i)} P_\pm^{(i)} + 2 s_i v_\pm^{(i)} \nonumber \\
     P_\pm^{(i)} &= - v_\pm^{(i)} - 2 v_\mp^{(i)} \frac{\partial E(\xi^{(+)} , \xi^{(-)})}{\partial \xi^{(\mp)}} 
 \end{align}
 with $\mathcal{P}^{(i)} = - (P^{(i)} + 2v^{(i)})$. The latter equation is the algebraic equation of motion for $v^{(N-i)}$, in case $s_i \neq 0$. Otherwise $v^{(i)}$ (and $v^{(N-i)}$) drop out of the Lagrangian completely. Substituting $K^{(i)}$, the first equation becomes:
 \begin{align}
     0 &= s_i D\star \mathcal{P}^{(i)}
     + \frac{1}{2} \sum_{j\neq i} \left( s_{i-j} [P^{(j)} , \star \mathcal{P}^{(i-j)}] + s_j  [P^{(j)} , \star \mathcal{P}^{(i-j)}] \right) \\
     &{} \quad + \frac{1}{2} \left( \sum_{j<i} (b_{i-j} - b_i + b_j) + \sum_{j>i} (b_{N+i-j} - b_i + b_j) \right) [P^{(j)},P^{(i-j)}]\, . \nonumber
 \end{align}
 For $s_i = 1$, $b_i = \left( 1 - \frac{2i}{N} \right)$, the `pure-spinor' type model, this reproduces the equations of motion in \eqref{eq:ZNdynamicaleom}. The second term $\frac{1}{2} \sum_{j\neq i} \left( [P^{(j)} , \star \mathcal{P}^{(i-j)}] + [P^{(j)} , \star \mathcal{P}^{(i-j)}] \right)$ vanishes due to the third constraint in \eqref{eq:Constraints}. As discussed above, full Hamiltonian integrability is shown by the fact that indeed the canonical momentum $\pi$ is essentially the field $\mathcal{P}_\tau$
 \begin{equation}
     \mathcal{P}_\tau^{a} = \pi^{a}.
 \end{equation}
The Lagrangian \eqref{eq:ZNdefaction} reproduces the more general deformation \eqref{eq:generalDef}, derived from the deformation of the Lax connection, with $f^{(\pm,i)} = - 2 s_i \frac{\partial}{\partial \xi^{(\mp)}} E(\xi^{(+)},\xi^{(-)})$. The original type of Lagrangian deformation in \cite{Ferko:2024ali} is included for $E(\xi^{(+)} , \xi^{(-)}) = E( \nu = \xi^{(+)} \xi^{(-)})$. It seems that the latter might be the only combination that is Lorentz-invariant, at least manifestly on the Lagrangian level. The dynamical \eqref{eq:ZNdynamicaleom} and algebraic \eqref{eq:Constraints} equations of motion take a (world-sheet) Lorentz-covariant form nonetheless. Similar to the principal chiral model and the symmetric space $\sigma$-model, the generalisation to $\xi^{(\pm)}_k \sim \text{Tr}(v^k_\pm)$ could be considered, as motivated from higher-spin current deformations in \cite{Bielli:2024ach}. For the symmetric space and the (Green--Schwarz) $\mathbb{Z}_4$-case, these deformations are discussed in \cite{Bielli:2024oif}.

\medskip
\newpage

\subsubsection*{Acknowledgements}

We are grateful to Christian Ferko and Liam Smith for useful correspondence, to  Benedikt K\"onig for collaboration during the initial stage of the project and to Stefan Theisen for helpful discussions. We also thank Daniele Bielli, Christian Ferko, Liam Smith and Gabriele Tartaglino-Mazzucchelli for agreeing to coordinate our preprints submissions. AK and DO wish to thank the University of Wroc\l aw and the Max Planck Institute for Gravitational Physics, Potsdam, for mutual warm hospitality during part of this work.  The research of DO is part of the project No. 2022/45/P/ST2/03995 co-funded by the National Science Centre and the European Union’s Horizon 2020 research and innovation programme under the Marie Sk\l odowska-Curie grant agreement no.~945339.

\vspace{10pt}
\includegraphics[width = 0.09 \textwidth]{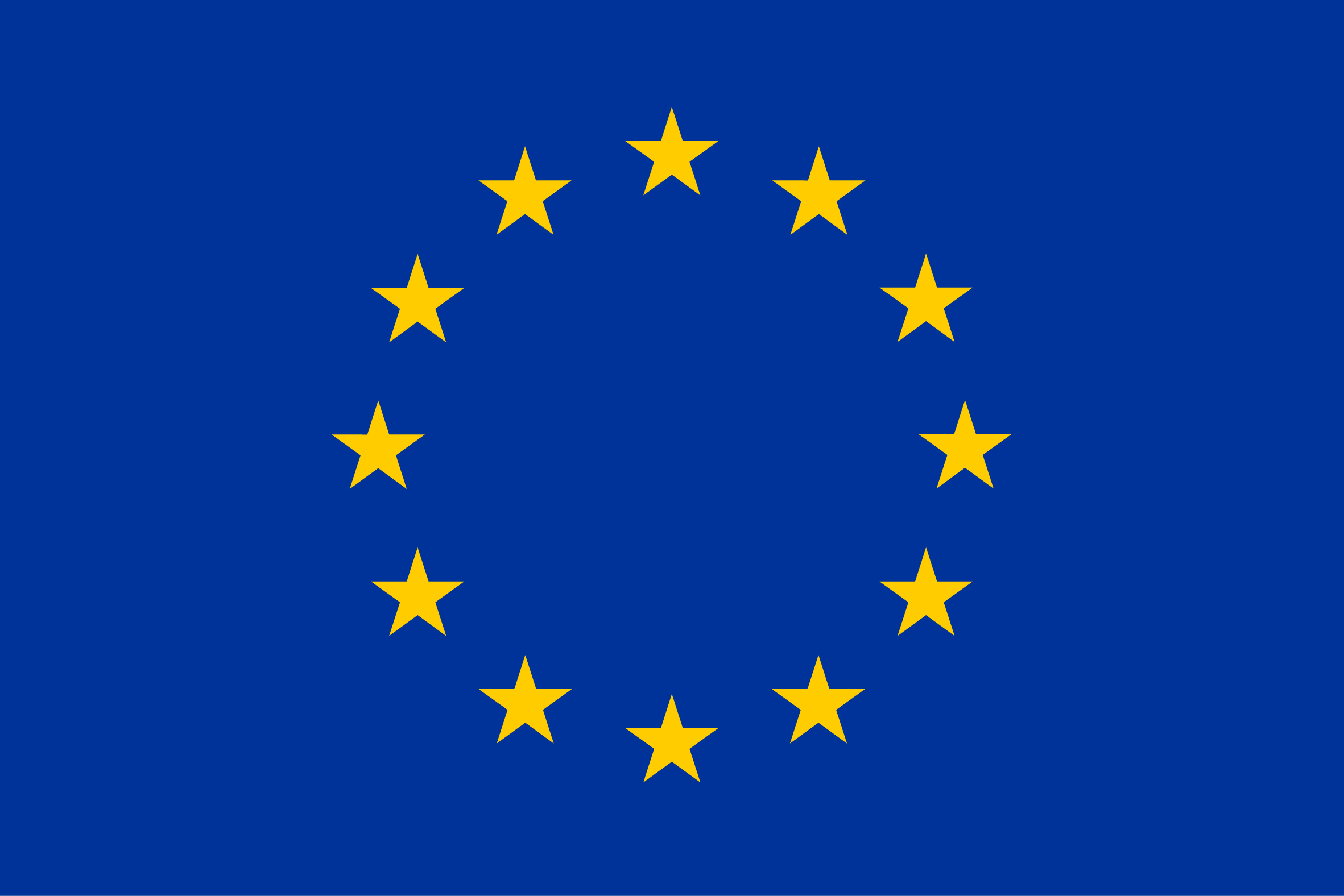} $\quad$
\includegraphics[width = 0.7 \textwidth]{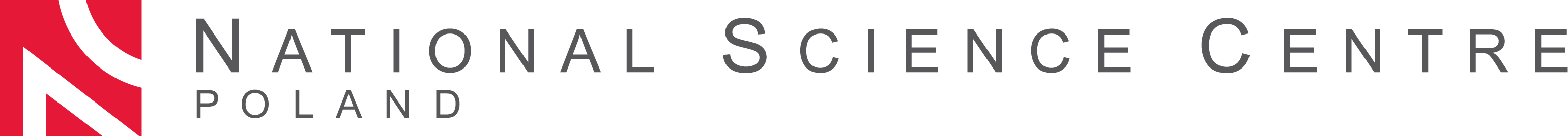}

\appendix
\section{Structure constants and anholonomy coefficients}
\label{sec:appA}

Here, we display the explicit form of the anholonomy coefficients for the symmetric space $G/K$~\eqref{eq:anhol} that are related to the structure constants of $\mathfrak{g}$ by $\Omega_{AB}{}^C=-f_{AB}{}^C$ and we split the indices according to $A=(a,\dot{a})$ etc., where $a$ refers to the coset directions $\mathfrak{p}=\mathfrak{g}^{(1)}$ and $\dot{a}$ to the isotropy directions $\mathfrak{k}=\mathfrak{g}^{(0)}$.
The derivatives are with respect to the coordinates $\varphi^m$ and $u^r$ introduced in section~\ref{sec:HamZ2}:
\begin{align}\label{eq:anholon}
\nonumber\Omega_{ab}{}^c&=0=E_a{}^mE_b{}^n\left(\partial_m E_n{}^c-\partial_n E_m{}^c\right)+\tilde{E}_a{}^r E_b{}^n\partial_rE_n{}^c-E_a{}^n \tilde{E}_b{}^r\partial_rE_n{}^c\, ,\\
\nonumber \Omega_{\dot{a}b}{}^c&=E_{\dot{a}}{}^sE_b{}^n\partial_s E_n{}^c\, ,\\
\nonumber \Omega_{ab}{}^{\dot{c}}&=E_a{}^mE_b{}^n\left(\partial_m E_n{}^{\dot{c}}-\partial_n E_m{}^{\dot{c}}\right)+\left(\tilde{E}_a{}^rE_b{}^n-E_a{}^n\tilde{E}_b{}^r\right)\partial_r E_n{}^{\dot{c}}+\tilde{E}_a{}^r\tilde{E}_b{}^s\left(\partial_r E_s{}^{\dot{c}}-\partial_s E_r{}^{\dot{c}}\right)\\
\Omega_{\dot{a}b}{}^{\dot{c}}&=0=E_{\dot{a}}{}^rE_b{}^m\partial_r E_m{}^{\dot{c}}+E_{\dot{a}}\tilde{E}_b{}^s\left(\partial_rE_s{}^{\dot{c}}-\partial_s E_r{}^{\dot{c}}\right)\, .
\end{align}



\providecommand{\href}[2]{#2}\begingroup\raggedright\endgroup

\end{document}